   \documentclass[fleqn,10pt,twocolumn]{wlscirep}
   \usepackage{amssymb}
   \usepackage[normalem]{ulem} 
\title{The impact of hexagonal boron nitride encapsulation on the structural and vibrational properties of few layer black phosphorus}
\author[1,*]{Magdalena Birowska}
\author[2]{Joanna Urban}
\author[2,3]{Michał Baranowski}
\author[2]{Duncan K. Maude}
\author[2]{Paulina Plochocka}
\author[1]{Nevill Gonzalez Szwacki}
\affil[1]{University of Warsaw, Faculty of Physics, Pasteura 5, 02-093 Warsaw, Poland}
\affil[2]{Laboratoire National des Champs Magnetiques Intenses, UPR 3228, CNRS-UGA-UPS-INSA, Grenoble and Toulouse, France}
\affil[3]{Department of Experimental Physics, Faculty of Fundamental Problems of Technology, Wroclaw University of Science and Technology, Wroclaw, Poland}

\affil[*]{Magdalena.Birowska@fuw.edu.pl}


\begin{document}
\begin{abstract}
The encapsulation of two-dimensional layered materials such as black phosphorus is of paramount importance for their stability in air. However, the encapsulation poses several questions, namely, how it affects, via the weak van der Waals forces, the properties of the black phosphorus and whether these properties can be tuned on demand. Prompted by these questions, we have investigated the impact of hexagonal boron nitride encapsulation on the structural and vibrational properties of few layer black phosphorus, using a first-principles method in the framework of density functional theory. We demonstrate that the encapsulation with hexagonal boron nitride imposes biaxial strain on the black phosphorus material, flattening its puckered structure, by decreasing the thickness of the layers via the increase of the puckered angle and the intra-layer P-P bonds. This work exemplifies the evolution of structural parameters in layered materials after the encapsulation process.
We find that after encapsulation, phosphorene (single layer black phosphorous) contracts by 1.1\% in the armchair direction and stretches by 1.3\% in the zigzag direction, whereas few layer black phosphorus mainly expands by up to 3\% in the armchair direction. However, these relatively small strains induced by the hexagonal BN, lead to significant changes in the vibrational properties of black phosphorus, with the redshifts of up to 10 cm$^{-1}$ of the high frequency optical mode $A_g^1$. In general, structural changes induced by the encapsulation process open the door to substrate controlled strain engineering in two-dimensional crystals.
\end{abstract}
\flushbottom
\maketitle
%
%

\section*{Introduction}

Two-dimensional (2D) materials offer novel physics and potential use in multiple applications \cite{tan2017recent, fiori2014electronics, xu2014spin, geim2013van,ling2015renaissance,carvalho2016phosphorene}. However, many of the layered materials are very sensitive to the local environment and ambient conditions\cite{zhou2007substrate, wang2008raman,calizo2007effect, tongay2013defects, tongay2013broad, zhang2013high,raja2017coulomb, stier2016probing}. Black phosphorus (BP) represents an extreme example of sensitivity to moisture and oxygen which can lead to catastrophic degradation on a time scale of only minutes \cite{favron2015photooxidation, castellanos2014isolation,koenig2014electric}. 
Therefore, to utilize its distinctive anisotropic properties in electronic, optoelectronic and thermo-electronic applications\cite{li2014black, fei2014enhanced, luo2015anisotropic,doi:10.1002/adfm.201502902, qiao2014high, wang2015highly,doi:10.1021/nn503893j,C5TC01809A}, methods of BP stabilization are unavoidable\cite{wood2014effective, na2014few}. Recent studies have shown that encapsulation with hexagonal boron nitride (hBN) protects BP from structural and chemical degradation, while improving its electrical properties \cite{doganov2015transport, cao2015quality, chen2015high, long2016achieving, avsar2015air} making hBN the most commonly used material for encapsulation. 

In contrast to other layered 2D materials such as transition metal dichalcogenides (TMDs) and graphene in BP the bonding between the consecutive layers is not purely of the van der Waals type\cite{C5NR06293D}. In BP atoms forms covalent bonds with three intralayer neighbors leaving a pair of lone electrons dangling out into the interlayer vacuum region \cite{doi:10.1021/acs.nanolett.5b00775, hu2016interlayer}. This results in enhanced interlayer interaction giving rise to a strong thickness dependence of the band gap \cite{li2017direct} and is also the main reason for the strong surface reactivity in air. In addition, it has been recently shown that in graphene-phosphorene heterostructures a large in-plane lattice contraction in phosphorene is induced, leading to a strong modification of the structural and electronic properties\cite{PhysRevMaterials.2.074001}.

It is therefore expected that the encapsulation of BP with hBN should affect its lattice structure. Since encapsulation is crucial to stabilize BP based devices, it is important to understand the interaction of BP with encapsulating materials. In addition, the detailed exploration of the structural and vibrational properties can expand our understanding on the lattice vibrations in this family of largely unexplored van der Waals heterostructures.  

In this paper we examine the influence of hBN encapsulation on the structural and vibrational properties of BP using density functional theory (DFT). We show that encapsulation strains the BP layer which has significant impact on the vibrational properties. Both non-encapsulated and encapsulated black phosphorus layers, exhibit anomalous evolution of phonon frequencies, which show a redshift with increasing number of layers. Encapsulation further enhances this redshift. Our studies shows that the underlying reason for this is the biaxial strain exerted on the hBN encapsulated BP layers which modifies the vibrational frequencies. The presented theoretical predictions are in good agreement with the results of Raman spectroscopy performed on hBN encapsulated BP. Our results shows that the presence of enhanced interlayer interaction in BP open the door to substrate controlled strain engineering in 2D crystals.

\section*{Methodology}
\subsection*{Experimental details}

The experimental results for hBN encapsulated BP presented in this work are based on data published and discussed in \cite{C7NR05588A}. All the details about the experimental setup and sample preparation can be also found in ref. \cite{C7NR05588A}.

\subsection*{Computational approach}
The calculations were performed in the framework of Density Functional Theory (DFT), within the local density approximation (LDA),\cite{PhysRevB.23.5048} as implemented in the Quantum Espresso package .\cite{0953-8984-21-39-395502} The electron-ion interaction was modeled using norm-conserving (NC) pseudopotentials.\cite{PhysRevB.40.2980} The kinetic energy cutoff for the plane-wave expansion of the pseudo-wave function was set to 60~Ry. A k-mesh of 15$\times$12$\times$1 was taken to sample the first Brillouin zone of the conventional unit cell for few layer BP. To do the slab calculations a 16~{\AA} of vacuum is added in order to avoid spurious interactions between replicates. The lattice parameters of each of the supercell were fully optimized and all of the atoms were relaxed until the maximal force per atom was less than $10^{-3}$ eV/\AA. With the optimized structures and self-consistent wave functions the phonon spectra and Raman intensities (non-resonant Raman coefficients using second-order response) were calculated as introduced by Lazzeri and Mauri.\cite{PhysRevLett.90.036401}  Density Functional Perturbation Theory (DFPT)\cite{RevModPhys.73.515} was employed to obtain phonon related properties. The threshold for the self-consistency of the DFPT was set to $10^{-14}$.

It was recently shown that the interlayer interaction between the layers of BP is not simply of the weak van der Waals type.\cite{doi:10.1021/acs.nanolett.5b00775} There exists sizable covalent interactions between the phosphorene layers significantly greater than for the other 2D materials such as graphene or MoS$_2$.\cite{doi:10.1021/acs.nanolett.5b00775}
In order to quantify the character of the interlayer bonding between the interface of hBN and BP layers, we calculate the interlayer binding energies as well as the interlayer equilibrium distances, for several structures; a bilayer of BP, a bilayer of graphene, hBN-BP, and hBN-graphene. All of the calculated values are collected in Table~\ref{bilayer}. We have employed two different exchange-correlation functionals, traditional LDA and optB86b-vdW\cite{PhysRevB.83.195131} which takes into account the van der Waals interactions. 
The energetics or the interlayer distances can be well reproduce using the appropriately chosen vdW exchange-correlation functional. The OptB86b is one of the several vdW functional proposed which are computationally cheaper than the highly computationally demanding RPA method\cite{PhysRevB.84.201401,PhysRevMaterials.2.034005}. The latter is considered to be one of the most accurate methods for describing the physics of vdW materials\cite{PDobson,PhysRevB.84.201401}.

\begin{table}\footnotesize
\caption{Equilibrium distances, $d$, and interlayer binding energies, $E_{b}$, for several structures. Positive values of energies indicate stable structures.} 
  \def\arraystretch{1.5}
  \label{bilayer}
  \begin{center}
  \begin{tabular}{  |c|  c|  c|  c| c |}
    \hline
    &   \multicolumn{2}{|c|}{ LDA} &   \multicolumn{2}{|c|}{OptB86b-vdW}  \\
     Structures& $d$ [\AA]& $E_{b}$ [meV/\AA$^2$] & $d$ [\AA]& $E_{b}$ [meV/\AA$^2$]\\
    \hline
   2L of BP& 2.90 & 26.04 & 3.12 & 30.57 \\
   2L of graphene  & 3.31 & 9.97 & 3.31 & 24.79 \\  
      hBN-BP   & 3.30 & 10.32 & 3.42 & 21.73\\  
      hBN-graphene  & 3.50 & 2.84 & 3.50 & 16.86 \\  
    \hline
  \end{tabular}
\end{center}
\end{table}

The interlayer distances  between the graphene and phosphorene layers (see Tab.~\ref{bilayer}) are reasonably well predicted by the LDA approach. In general the LDA approach correctly described the chemical covalent bonds, however, for fortuitous reasons it can also reproduce quite well the interlayer distances in layered materials.\cite{Acta2011} Similar results have been previously obtained by using different codes \cite{Acta2011} and van der Waals type of functional.\cite{doi:10.1021/acs.nanolett.5b00775} The interaction between the graphene layers are strongly underestimated by the LDA approach, the predicted interlayer binding energy is 9.97  meV/\AA$^2$ compared to  the correct value of 24.79 meV/\AA$^2$ when van der Waals interactions are taken into account. In case of two layer BP the LDA and optB86b-vdW approaches give similar values, of 26.04 meV/\AA$^2$ and 30.57 meV/\AA$^2$, respectively, indicating the rather quasi covalent character of the bonding.\cite{doi:10.1021/acs.nanolett.5b00775} This suggests that the LDA approach should capture most of the relevant physics of the interaction between the BP layers. In contrast, the LDA calculations clearly underestimate the binding energy for two layer graphene, hBN-BP and hBN-graphene showing the importance of the van der Waals interaction in these system. We note that  a similar value 21.6~meV/\AA$^2$ for hBN-BP \cite{PhysRevB.91.115413} has been previously obtained using a different type of vdW exchange-correlation functional.

For our detailed calculations on hBN encapsulated few layer BP we have decided to use the LDA approach which provides the best description of the interlayer phonon frequencies.\cite{doi:10.1021/acs.nanolett.5b00775} The encapsulation process in our calculations is captured by introducing biaxial strain directly calculated for the hBN-BP-hBN heterostructures. Due to the weak van der Waals forces we expect no changes in electronic structures\cite{Acta2011} or negligible charge transfer at the interface\cite{doi:10.1021/acs.jpcc.5b02634}, which justifies our approach. We use the smaller $(1\times1)$ lateral supercell of BP layers, where the strains are respectively rescaled. The BP strained supercells, are then used to calculate the vibrational properties.

\subsection*{Structural properties of freestanding and encapsulated BP layers}
\begin{figure}[h]
 \centering
 \includegraphics[width=0.45\textwidth]{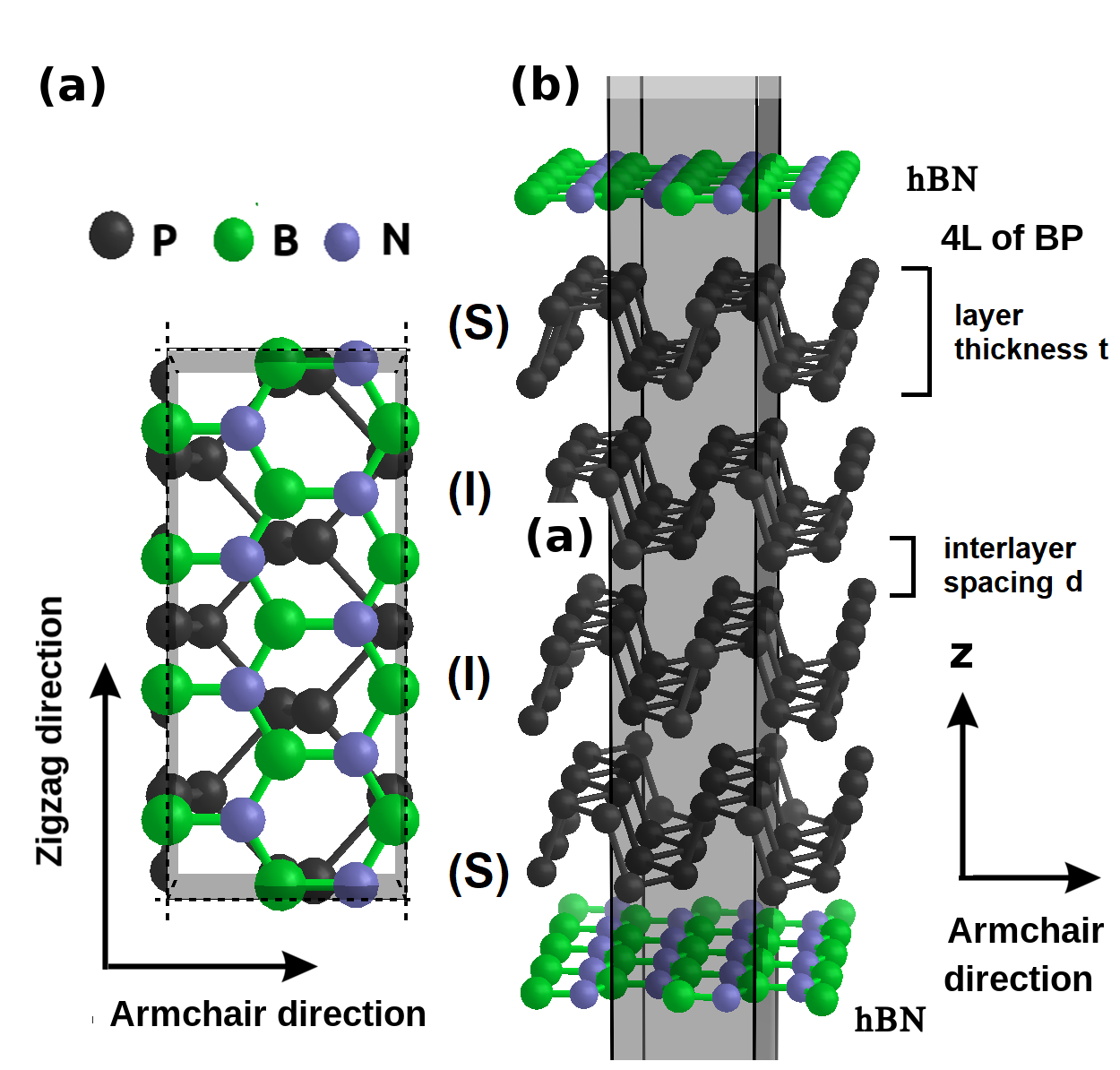}
 \caption{\label{supercell} The supercell of the hBN encapsulated four layer BP. The supercell is constructed by combining the orthogonal supercell of hBN $(\sqrt{3}\times1)$ and $(3\times1)$ supercell of phosphorene. The (a) lateral and (b) side views of the supercell are presented. The S and I letters denote the surface and inner layers, respectively.}
\end{figure}
\begin{figure*}[h!]
\centering
\includegraphics[width=0.7\textwidth]{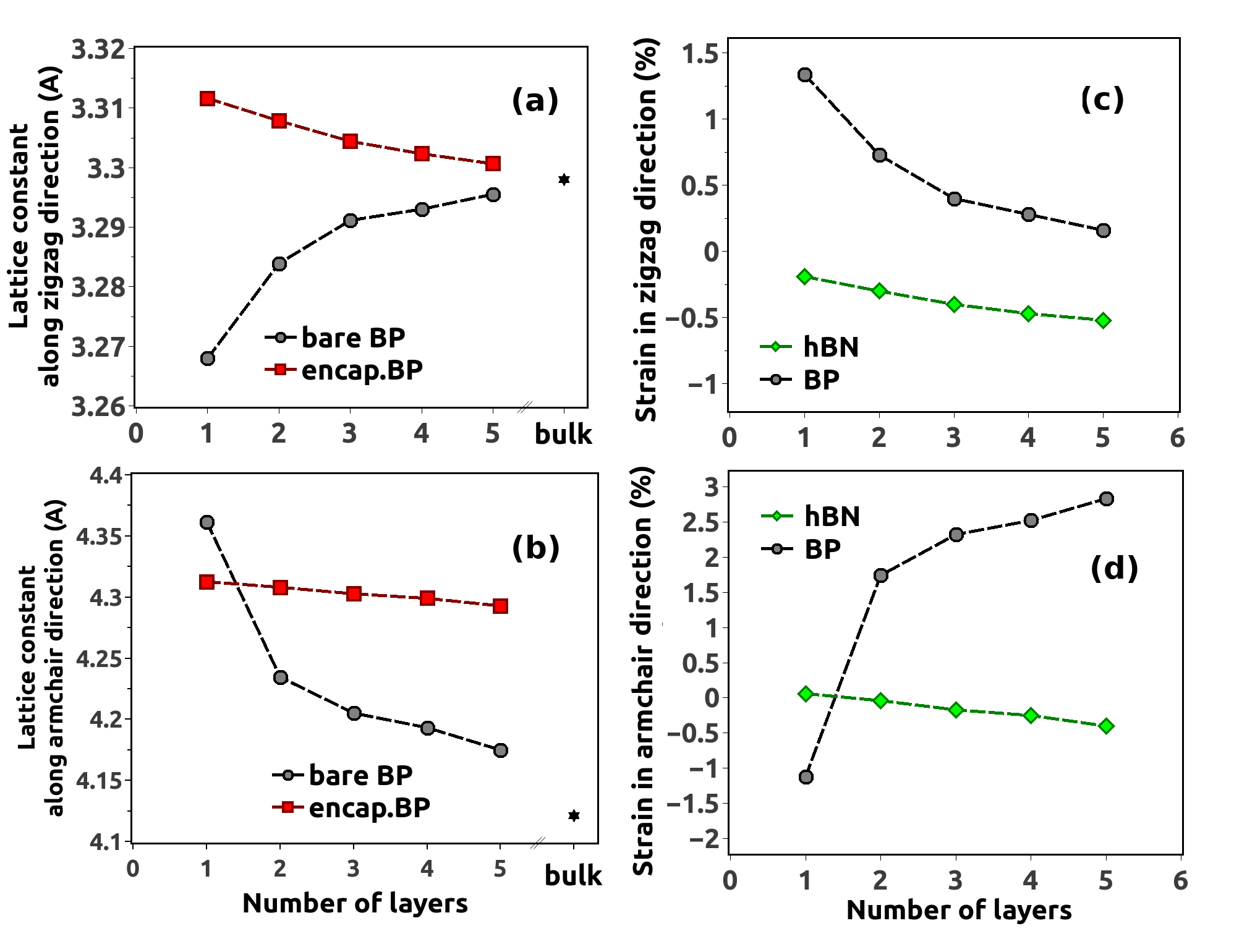}
\caption{\label{str} Dependence of lattice parameters on the number of BP layers (on the left) and the strain effect in the encapsulated layers (on the right) presented for the in-plane directions. In the (c) and (d) panels the strains are presented for the BP layers (black curve) and hBN layer (green curve) in the encapsulated  BP structures. A large stretchability along the armchair direction of the hBN encapsulated BP layers is observed.}
\end{figure*}
The epitaxy of hBN/BP/hBN imposes the hexagonal structure of BN on the orthogonal superlattice of BP layers. This is in contrast to widely examined 2D heterostrcutures  such as graphene/TMD, graphene/BN, graphene/silicene, TMD/TMD, where the constituent layers possess the same hexagonal lattice structure. Therefore, each supercell is constructed by combining  an orthogonal $(\sqrt{3}\times1)$ supercell of hBN and $(3\times1)$ supercell of phosphorene (see Fig.~\ref{supercell}).\cite{doi:10.1021/acs.jpcc.5b02634} 
Note, that there is a lack of experimental data concerning the arrangements of the hBN and BP layers. Thus, it is possible, that the mutual alignment of the hBN and BP layers in the chosen supercell differs from reality.
 
In Figs. \ref{str}(a) and \ref{str}(b), we present the results of lattice parameters for the bare BP layers (black curve) and encapsulated BP layers\footnote{Encapsulation here means that we have as a hBN layer above and below the BP slab. See Fig.~\ref{supercell} in Suppl. Materials.} (red curve).
Our results for the bare structures show clear trends of the in-plane lattice parameters, namely, a large decrease in armchair lattice constant by about 0.20~{\AA} with increasing number of layers, accompanied by a much smaller increase of 0.03~{\AA} of the zigzag lattice constant. Similar trends were  previously obtained for bare samples with the vdW approach.\cite{C5NR06293D} In contrast the encapsulated BP layers show a small decrease in the lattice parameters as the function of the number of the layers for both in-plane directions, indicating that the BP layers are  significantly strained by the hBN encapsulation.

\begin{figure*}
\centering
\includegraphics[width=0.95\textwidth]{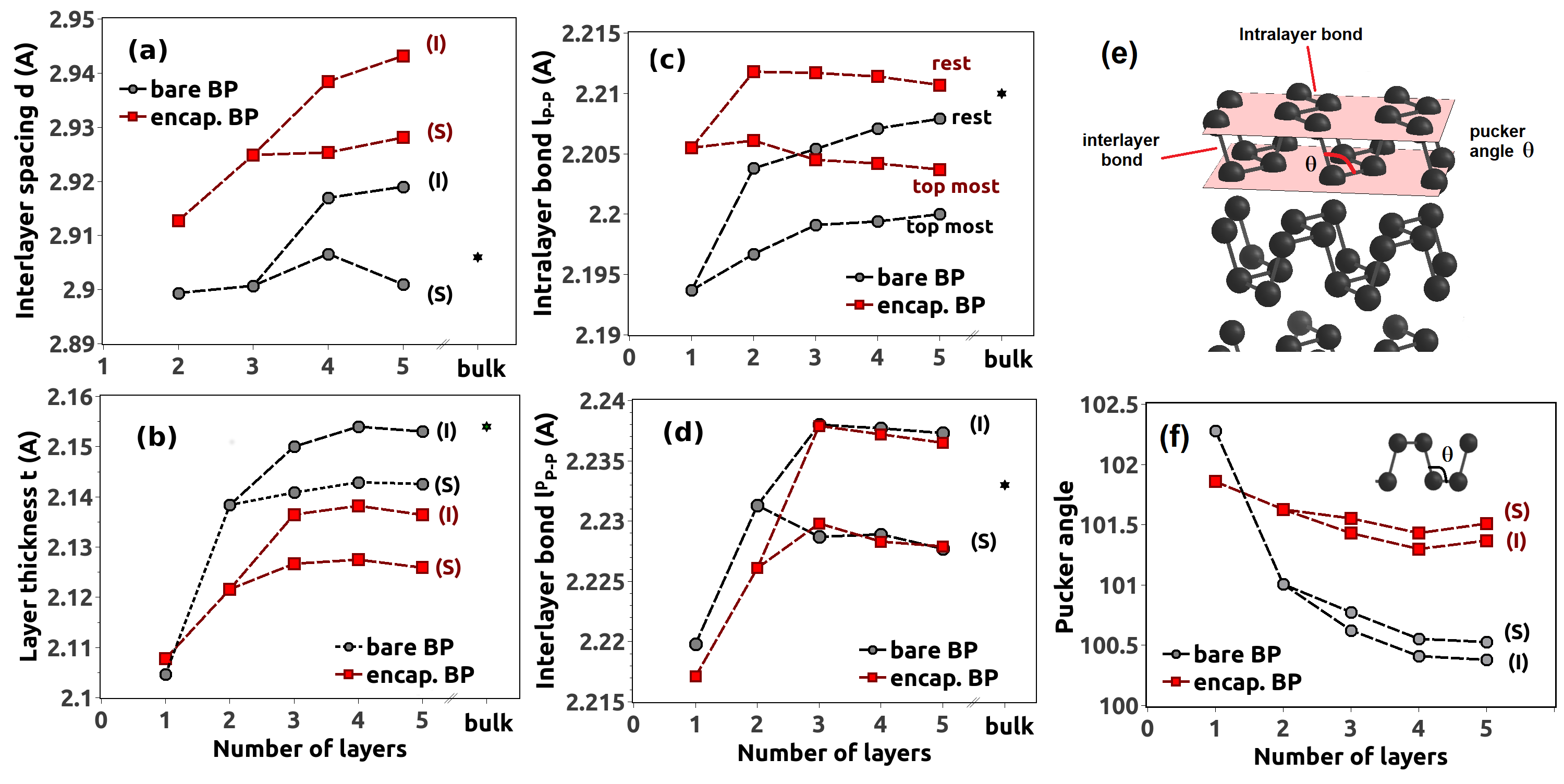}
\caption{\label{sp} The comparison of the structural parameters of the freestanding BP layers with hBN encapsulated BP layers. The structural parameters: interlayer spacing $d$ (a), layer thickness $t$ (b), length of the intralayer bond $l_{P-P}$ (c), length of the interlayer bond $l^p_{P-P}$ (d), and the pucker angle $\theta$. All of them exhibit surface-inner splitting. The (I) and (S) marks denote the atoms from the surface and inner layers, respectively. The top most in (c) indicates the intralayer P-P bond from the atoms in the layer next to the vacuum region, whereas the rest denotes the largest value of the intralayer bond taken from layers below the surface. }
\end{figure*}

In order to quantify this effect, we calculate the biaxial strains for both hBN encapsulated BP layers and the hBN layer itself (see Figs. \ref{str}(c) and \ref{str}(d)).
We define the uniaxial strain along zigzag (or armchair) direction for a given number $nL$ of  BP layers as 
\begin{equation}
\epsilon_{BP} =\frac{a_{n}^{e-BP} - a_{n}^{b-BP}}{a_{n}^{b-BP}} \times 100\%,\nonumber
\end{equation}
and for the hBN encapsulated layer as 
\begin{equation}
\epsilon_{hBN} =\frac{a_{n}^{hBN}-a_{0}^{hBN}}{a_{0}^{hBN}} \times 100\%,  \nonumber
\end{equation}
where $a_{n}^{e-BP}$ and $a_{n}^{b-BP}$ denote the optimized lattice constants of encapsulated (e-BP) and bare (b-BP) optimized n-layer BP structures, whereas $a_{n}^{hBN}$ indicates a single layer of hBN used to encapsulate the the $n$ BP layers ($n=0$ corresponds to bare hBN).  Positive and negative values of $\epsilon$ correspond to tensile and compressive strains.

Figures \ref{str}(c) and \ref{str}(d) show the strain of the examined structures as a function of the number of BP layers. The capping hBN layer is weakly compressed, and this compression slightly increases with the number of BP layers up to 0.52\% and 0.4\% for the zigzag and armchair directions, respectively. In striking contrast to the capping layer, the BP layers are tightly stretched (except for single layer BP in the armchair direction) reaching value of 2.83\% and 1.34\%, for 5 layer BP in armchair and single layer BP in the zigzag directions, respectively. These results reveal the well known fact that hBN is a very stiff material, and hence, it enforces its own lattice structure on the entire vdW  heterostructure. 

We now calculate how the encapsulation via the induced lattice strain changes the structural parameters of the constituent BP layers. The results are illustrated in Fig.~\ref{sp}, where the black curves represent the freestanding BP layers, and red curves, the encapsulated ones. Under the encapsulation process the thickness of the BP layer (see Fig.~\ref{sp}(b)) 
decreases by about 0.015~\AA, whereas the length of the in-plane intralayer bond increases by approximately 0.007~\AA (see Fig.~\ref{sp}(c)) which is accompanied by an increase of the pucker angle $\theta$ of $\simeq 0.8^{\circ}$  (see Fig.~\ref{sp}(f)). At the same time, the length of the interlayer bond remains almost unchanged (see Fig.~\ref{sp}(d)). These  results clearly indicate that the encapsulation process flattens the puckered structure of the BP layers via an the expansion of the intralayer bonds (see Fig.~\ref{sp}(c)). Moreover, in all of the structural parameters a splitting into two branches is observed, which directly distinguish the atoms on the surface (lower branch) and the atoms in the inner layers (upper branch).

In addition, an extraordinary stretchability of BP layers along the armchair direction is observed, with opposite increasing trend to the zigzag direction. The reason for this effect lies in the puckered honeycomb structure of the BP layers. Our and previously reported results\cite{APL2014Wei} demonstrate that  is relatively  easy to flatten the BP structure by increasing the pucker angle.  The difference in strains along in plane along the armchair and zigzag directions reflects the strong anisotropic properties of BP, which might  be potentially useful in strain-engineering applications.

\subsection*{Vibrational properties of bare and encapsulated BP layers}
   \begin{figure*}
 \centering
 \includegraphics[width=0.7\textwidth]{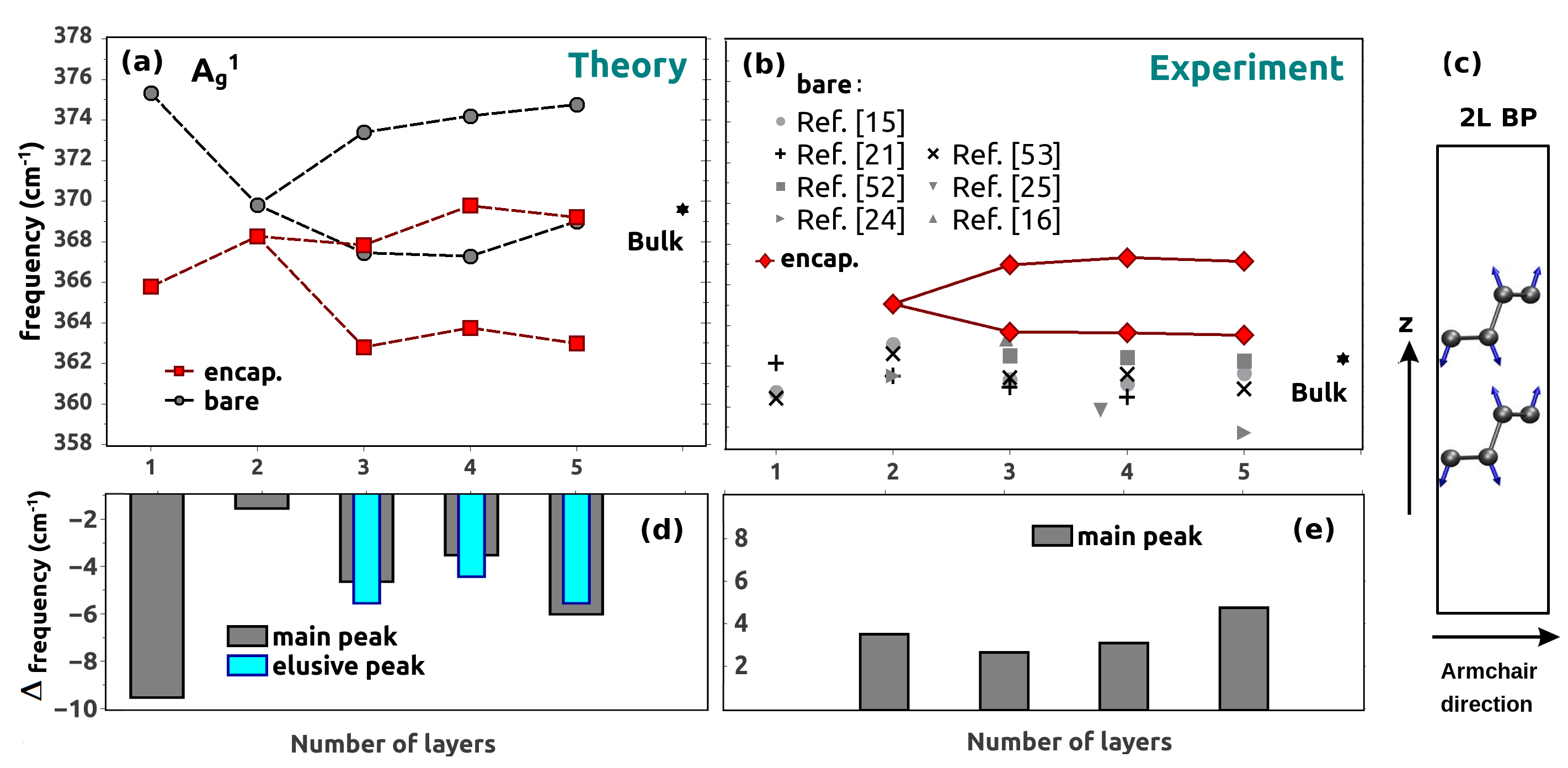}
 \caption{\label{ag1tot} A comparison between the experimental (b), (e) and theoretical results (a), (d) for bare (black curves, grey scale points) and encapsulated BP (red curves) evolution of $A_g^1$ optical mode at $\Gamma$ point as function of the number of BP layers. The experimental data for bare samples are taken from refs. \cite{favron2015photooxidation,castellanos2014isolation,doi:10.1002/adfm.201502902,doi:10.1021/nn503893j,C5TC01809A,Lu2014,cPh}. A splitting of the frequencies is clearly visible for experimental (encapsulated layers) and theoretical results. The frequency shift $\Delta$ of the  $A_g^1$ mode upon encapsulation is presented in (d) and (e). This shift is defined as $\Delta=f_{b,n}^{encap.}-f_{b,n}^{bare}$, where $b$ ($=l,u$) indicates  lower branch $l$  (main peak) or upper branch $u$  (elusive peak) of frequencies for a given number $n$ of BP layers. The negative and positive values of this shift indicate a redshift or blueshift in frequency. In case of the experimental results, only the results for the main peaks are shown ($b=l$) in (d), where $f_{l,n}^{bare}$ indicates the lowest frequency of all bare results presented in (b) for a given number of layers. To facilitate a comparison of the results, all of the theoretical frequencies shown in (a) have been moved up by 10 cm$^{-1}$, to match the experimental values. (c) Schematic diagram  for vibrational $A_g^1$ mode for 2L of BP is presented.
 }
  \end{figure*}
  
  Now we will present results, which show that the structural changes can significantly affect the vibrational properties of the vdW heterostructures. Although the vibrational properties of 2D structures such as graphene, transition metal dichalgonides (TMDs) or black phosphorus (BP) are well studied, the vibrational properties of vdW heterostructures remain largely unexplored. Here we focus on the A$_g^1$ optical intralayer mode for which encapsulation produces the most pronounced vibrational changes observed experimentally\cite{C7NR05588A}. For the other intralayer modes such as A$_g^2$ and B$_{2g}$ the calculated frequencies do not significantly change upon encapsulation. The intralayer modes  involve the vibrations from the in-plane chemical bonding, and thus, can give insight into these bonds.\cite{doi:10.1021/acs.nanolett.5b01117} 
  
  Here we present the results of the hBN encapsulation as a function of the number of BP layers, by taking into account the biaxial strains resulting from hBN capping layers. This aspect is the main difference between our and previously reported results under the strain.\cite{APL2016Tokar,APL2014Fei,C5NR06293D} In previous studies concerning vibrational properties, mostly the phonon frequencies were examined. In this work we also present the theoretical predictions and experimental results for the non-resonant Raman intensities.

 Recently it has been shown that the evolution of the frequency modes is anomalous, namely a redshift in frequency is observed with an increasing number of layers,\cite{Lu2014,C7NR05588A} in contrast to weakly coupled 2D materials where a blueshift is predicted.\cite{doi:10.1021/nn1003937} An obvious question is can this anomalous behaviour in encapsulated layers be easily understood in terms of the structural changes, as suggested in Ref.\cite{C5NR06293D} or if a different mechanism has to be considered.

All of the results presented here are calculated for $(1\times1)$ supercell, and the encapsulation effect is included by taking into account the biaxial strain introduced by the presence of the hBN layers. The strain depends strongly on the thickness of the examined vdW heterostructures. Fig.~\ref{ag1tot}  shows the evolution of the $A_g^1$ high frequency mode for theoretical and experimental results. Both, theory and experiment results, exhibit similar trends, namely two branches (a splitting) are observed for three layers and above. 

The frequency splitting originates from the different stiffness of the surface and inner layers. The surface layer is stiffer in comparison to inner layers for both bare and encapsulated structures, which is revealed by the different structural parameters of surface atoms compared to atoms in the inner layers. Namely, intralayer and interlayer bonds (see Fig.~\ref{sp}(c) and (d)), the interlayer spacing (Fig.~\ref{sp}(a)) and the layer thickness (Fig.~\ref{sp}(b)) are all reduced for atoms at the surface. This reflects the stronger attraction between atoms from surface layers than between atoms from inner layers, independently of whether we are dealing with bare or encapsulated BP layers.

 In addition, the encapsulation process can significantly decreases the frequency of the studied modes, and this decrease strongly depends on the thickness of the structures reaching a value of 10\,cm$^{-1}$ for single layer BP and minimum value of 2\,cm$^{-1}$ for two layer BP in comparison to freestanding BP layers (see Fig.~\ref{ag1tot}(d)). The origin of this redshift stems from the increase of the interlayer spacing (red curve in comparison to black curve in Fig.~\ref{sp}(a)), which results in smaller attraction between the layers, as well as the lateral extension of the puckered structure of BP layers via the increased pucker angle (see Fig.~\ref{sp}(f)) and the elongation of the intralayer bonds (see Fig.~\ref{sp}(c)).

 \subsection*{Experiment versus theory}
 
Both, theoretical as well as the experimental results, for encapsulated layers exhibit similar trends for the evolution of the $A_g^1$ mode (see Fig.~\ref{ag1tot}), namely two branches are observed for three layers and above. The Raman measurements (red curve in Fig.~\ref{ag1tot}(b)) were carried out for encapsulated BP layers.
No splitting for non-encapsulated black phosphorus has been reported, in contrast to to theoretical results, which also predict splitting for bare phosphorus. This can be understood as a result of material quality improvement after encapsulation and narrowing of the observed modes \cite{C7NR05588A} which facilitates detection of the elusive peak.

\begin{figure}[h!]
 \centering
 \includegraphics[width=0.4\textwidth]{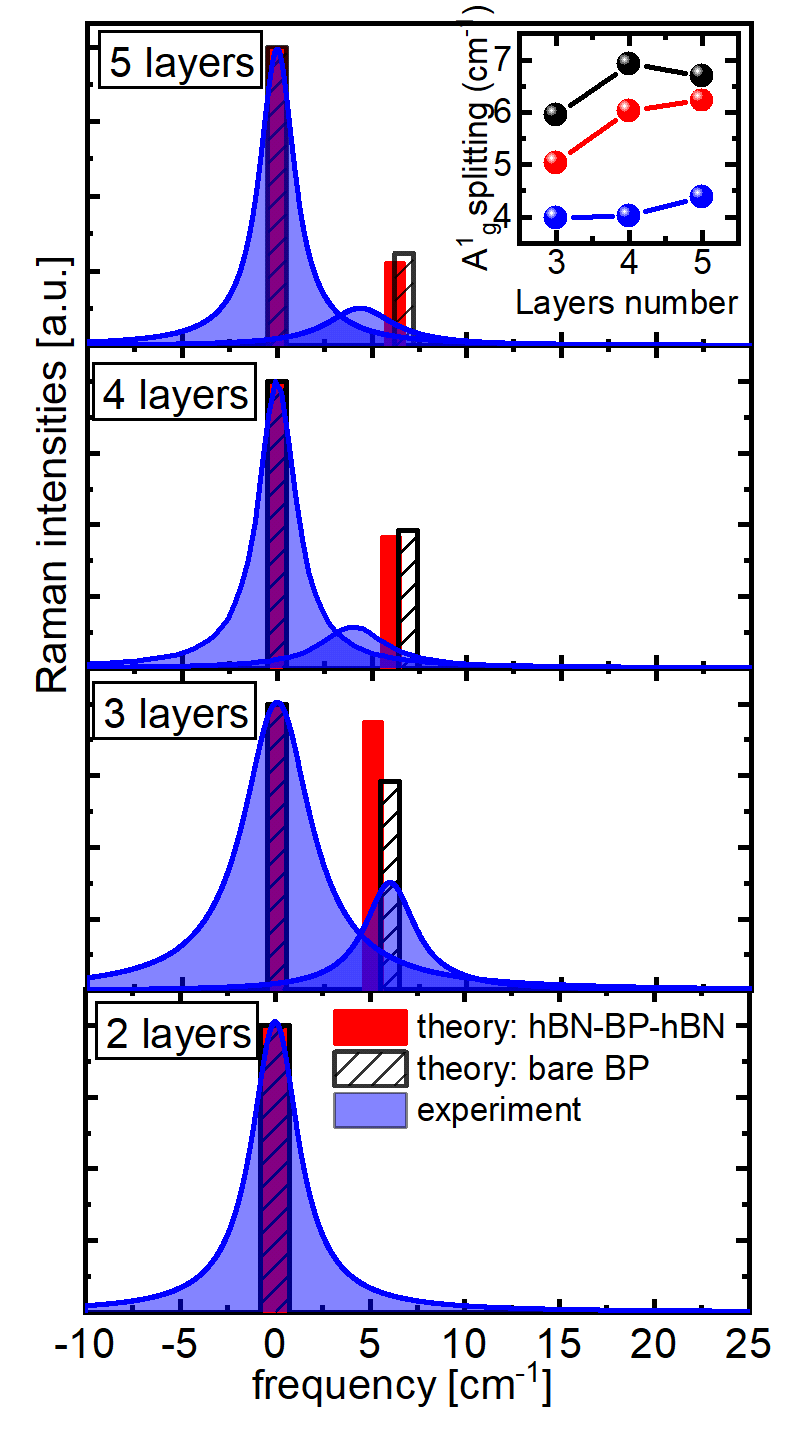}
 \caption{\label{raman} Raman intensity of $A_g^1$ mode and its splitting (inset), for both bare and  encapsulated BP layers. The main peak (lower frequency branch in Fig.~\ref{ag1tot}) and the elusive peak (higher frequency branch in Fig.~\ref{ag1tot}) as a function of the number of layers are presented. For a comparison of the  theoretical and experimental results, the frequency of the main peak is arbitrarily set to zero. The splitting is defined as the difference in frequency of the elusive peak and the main peak.}
 \end{figure}

In addition, both the experimental and theoretical results show that the most pronounced elusive peak\footnote{The main and elusive peaks are lower and higher frequency branches shown in Fig.~\ref{ag1tot}, respectively.} is observed for three layers (3L), and its intensity decreases with the number of the BP layers in the system (see Fig.~\ref{raman}). This effect can be understood in terms of the relative number of atoms at the surface (origin of elusive peak), compared to the number of atoms present in inner layers (origin of the main peak), as the  Raman intensity is proportional to the number of atoms in the system. In case of 3 layers, there exists the largest ratio for the number of surface atoms to number of inner atoms. Therefore, the elusive peak is quite strong for 3 layers, and its intensity monotonically decreases with the increase of the layers in the systems.  Moreover, for the most pronounced case (3L), the elusive peak can be even slightly enhanced by encapsulation process (see Fig.~\ref{raman}), and from the Raman intensity point of view, it can be easier to detect in experiment.

Surprisingly the theoretical results shows that after encapsulation the energy of the A$_g^1$ mode decreases for all number of layers. This is in contrast to experimental results where hBN encapsulation leads to an A$_g^1$ frequency which is higher than in non-encapsulated samples (see Fig.~\ref{ag1tot} (b),(e)). However, non-encapsulated black phosphorus is not actually freestanding as in all reports BP structure are placed on some kind of substrate, usually on SiO$_2$. The SiO$_2$ is much more ``rugged'' than hBN and posses a large number of dangling bonds so it is expected that its interaction with BP is stronger than hBN increasing the value of the strain redshifting the  A$_g^1$ mode. It is worth to note that this argumentation leads to the conclusion that all of the so far reported studies have been in fact performed on the strained BP. Hence the further experimental and theoretical studies are highly needed and desirable in this aspect.

\section*{Conclusions}

In conclusion we have performed detailed theoretical studies of the vibrational properties of bare and hBN encapsulated with BP. We have shown, that the strong interaction between the individual layers of BP results in the different stiffness of the surface and inner layers. This is directly reflected by the splitting of the $A_g^1$ Raman mode for three layer and above BP. Furthermore, we have demonstrated that the strain of the BP is strongly affected by the encapsulating layer. In particular we show that the commonly used hBN encapsulation has a significant impact on the vibrational properties of the BP layers. Namely, the $A_g^1$ interlayer mode frequencies decrease by $\simeq (2-10)$\,cm$^{-1}$ compared to bare BP (depending on the number of layers). This redshift results from the smaller attraction between the layers as well as and flattening of the puckered structure of BP via the extension of the intralayer bonds and the pucker angle. Our theoretical prediction are in good quantitative agreement with experimental results. Importantly, our calculations reveal that in case of BP even van der Waals interaction of BP with encapsulating material can significantly affects it properties due to strain formation. This interaction can be become even more important in case of less ``flat'' substrate and the influence of the encapsulation layer and cannot be neglected. The result of this study provide a starting point to define a strategy for the strain engineering in BP based heterostructures.

\section*{Acknowledgements}
This work is funded by the National Science Centre grant no. UMO-
2016/23/D/ST3/03446. Access to computing facilities of PL-Grid Polish Infrastructure for Supporting Computational Science
in the European Research Space and of the Interdisciplinary Center of Modeling (ICM), University of Warsaw
is gratefully acknowledged. The experimental work was partially supported by BLAPHENE and STRABOT projects, which received funding from the IDEX Toulouse, Emergence program,  ``Programme des Investissements d'Avenir'' under the program ANR-11-IDEX-0002-02, reference ANR-10-LABX-0037-NEXT, and
by the PAN--CNRS collaboration within the PICS 2016-2018 agreement. We thank EPSRC for funding through grant EP/M05173/1. Micha{\l} Baranowski appreciates support from the Polish Ministry of Science and Higher Education  within  the  Mobilnosc  Plus program (grant no. 1648/MOB/V/2017/0). N.G.Sz. gratefully acknowledges the support of the National Science Centre through grant no. UMO-2016/23/B/ST3/03575.







\bibliography{sample}


\end{document}